\documentclass{PoS}

\title{Event-by-event Particle Yield Ratio Fluctuations in the CBM and NA49 Experiments}

\ShortTitle{Particle Ratio Fluctuations in CBM and NA49}

\author{\speaker{Dmytro Kresan} for the CBM and NA49 collaborations \\
        GSI, Darmstadt\\
        E-mail: \email{D.Kresan@gsi.de}}

\abstract{Non-statistical event-by-event fluctuations are considered
an important signal for the critical endpoint of the QCD phase 
diagram. Event-by-event fluctuations of different observables are
thus investigated in detail in current experiments.
In this contribution, a study of the
centrality dependence of event-by-event fluctuations of particle
yield ratios by the NA49 experiment in Pb+Pb collisions at 158$A$~GeV
beam energy is presented for the first time. An increase of the absolute
values of dynamical fluctuations towards lower centralities is observed.
The influence of resonance decays will be discussed. \\
Event-by-event fluctuations are an important observable to be studied
at the future CBM experiment at FAIR. CBM will investigate the
intermediate region of the QCD phase diagram in great detail searching
for the first order phase transition line and the expected critical
endpoint. It is therefore important to investigate the sensitivity of
the CBM detector to particle ratio fluctuations in Au+Au collisions at
10-45~AGeV beam energy. Detailed simulation studies will be presented.}

\FullConference{5th International Workshop on Critical Point and Onset of Deconfinement - CPOD 2009,\\
		 June 08 - 12 2009\\
		 Brookhaven National Laboratory, Long Island, New York, USA}

\begin{document}

\section{Introduction}

\begin{figure}[ht]
\center
\includegraphics[width=0.5\textwidth]{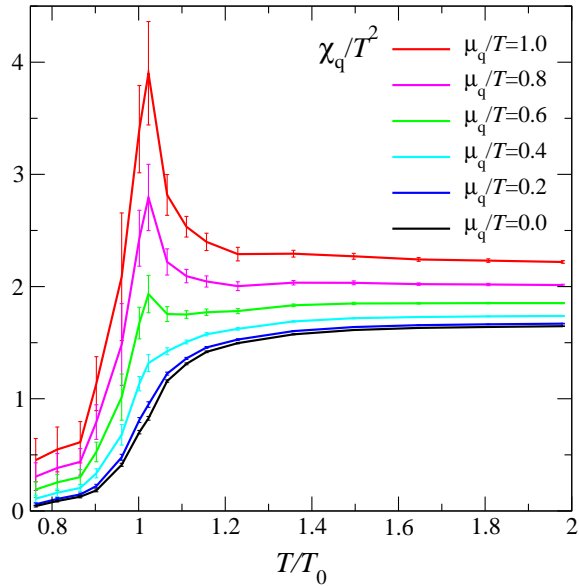}
\caption{The quark number susceptibility as a function of temperature
for different quark chemical potentials~\cite{Allton:2003vx}.}
\label{fig:intro:chiqT}
\end{figure}

Recent lattice QCD calculations expect a first order phase transition
from hadronic to partonic degrees of freedom at finite temperature and
baryon chemical potential~\cite{Fodor:2004nz,Aoki:2004iq}. This first order
transition line ends with a critical endpoint~\cite{Ejiri:2007ga}. The
search for either the first order phase transition or the critical endpoint
is a challenging task in modern high energy heavy-ion physics. \\
Due to density fluctuations at the critical point or in the coexistance
region at a first order phase transition, fluctuations in particle yields
and kinematic properties may occur. Indeed, in lattice QCD calculations
the quark number susceptibility develops a peak at the critical temperature
when increasing the quark chemical potential and thus approaching the
expected critical point (figure~\ref{fig:intro:chiqT})~\cite{Allton:2003vx}.
Fluctuations as a measure of susceptibilities~\cite{Koch:2008ia} should
than be enhanced.

\begin{figure}[ht]
\center
\includegraphics[width=0.5\textwidth]{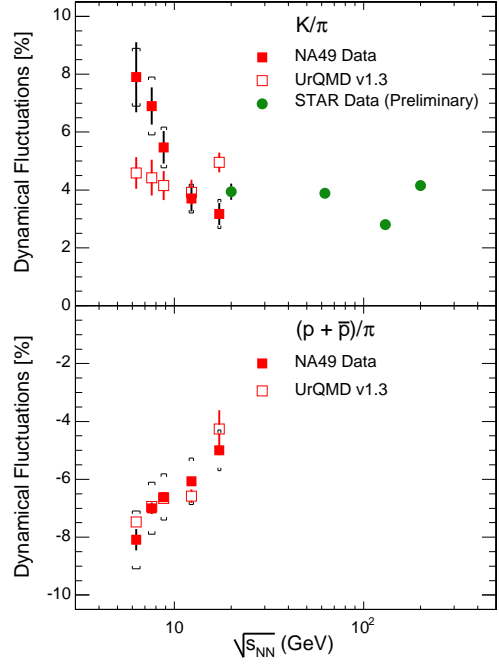}
\caption{Energy dependence of the dynamical fluctuations of the K/$\pi$
and p/$\pi$ ratios in central A + A collisions published by the
NA49 and STAR experiments~\cite{:2008ca,:2009if}.}
\label{fig:intro:ebefluct}
\end{figure}

In this work we will concentrate on event-by-event fluctuations
of particle yield ratios, such as kaon to pion (K/$\pi$), proton to pion
(p/$\pi$) and kaon to proton (K/p) ratios. Recently, the NA49 and STAR
collaborations have published their results on the energy dependence of
the dynamical fluctuations of K/$\pi$ and p/$\pi$ ratios~\cite{:2008ca,:2009if}.
The measured dependence is illustrated in figure~\ref{fig:intro:ebefluct}.

The dynamical fluctuations of the K/$\pi$ ratio increase towards lower collision
energies and are not reproduced by the UrQMD model, while the p/$\pi$ ratio fluctuations
from the data analysis are in good agreement with model calculations. \\
In~\cite{:2008ca} it was discussed, that this increase of the K/$\pi$
ratio fluctuations towards lower beam energies could be associated with a decrease
of particle multiplicities. Thus one would like to study the centrality dependence
of the fluctuation signal in order to understand the behaviour of this observable
with the changing average multiplicity of the accepted tracks. In addition,
it is neccessary to re-measure the low energy region with better experimental
resolution and larger acceptance to provide high quality data on this
observable, which is one of the goals of the future CBM experiment.

\section{Centrality Dependence of Particle Ratio Fluctuations in NA49}

NA49 is a fixed target experiment at the CERN SPS. Measurements of p + p, p + A
and A + A collisions were recordered at different energies. NA49 is a large
acceptance hadron spectrometer which includes two dipole magnets for momentum
determination, two vertex TPCs for the measurement of vertices and particle identification
(PID) using $dE/dx$, two main TPCs for PID with $dE/dx$, two TOF walls for PID at
midrapidity using time-of-flight and a Zero Degree Calorimeter for centrality
determination. A detailed description of the NA49 detector can be found
in~\cite{Afanasev:1999iu}. \\
In the current analysis, the specific energy loss $dE/dx$, measured by the main TPCs
of NA49 was used for particle identification. The NA49 spectrometer works in the
relativistic rise region of the Bethe Bloch parametrisation of
$dE/dx$~\cite{Wenig:1998vv}. This implies some restrictions on the PID performance,
namely a lower momentum cut of 3~GeV/c and no possibility of PID on the track-by-track
level. Thus, particles have to be identified on a statistical basis by extracting
the relative yields in a single event using the Maximum Likelihood Method. The data
analysis technique is described in detail in~\cite{bib:croland,bib:dkresan}. \\
In this section we will focus on the measured centrality dependence of the dynamical
fluctuations of particle yield ratios. The centrality of an A + A collision was
determined by measuring the total energy of the projectile spectators deposited
in the Zero Degree Calorimeter of NA49.
The dynamical fluctuations of a particle yield ratio are defined as the geometrical difference
between the relative width ($\sigma$=RMS/MEAN) of the eventwise ratio distribution for
data ($\sigma_{data}$) and mixed ($\sigma_{mix}$) events:

\begin{equation}
\sigma_{dyn} = sign(\sigma_{data}-\sigma_{mix})\sqrt{\Big|\sigma_{data}^{2}-\sigma_{mix}^{2}\Big|}
\label{eq:na49:sdyn}
\end{equation}

\subsection{Centrality Bin Size}

\begin{figure}[ht]
\includegraphics[width=0.49\textwidth]{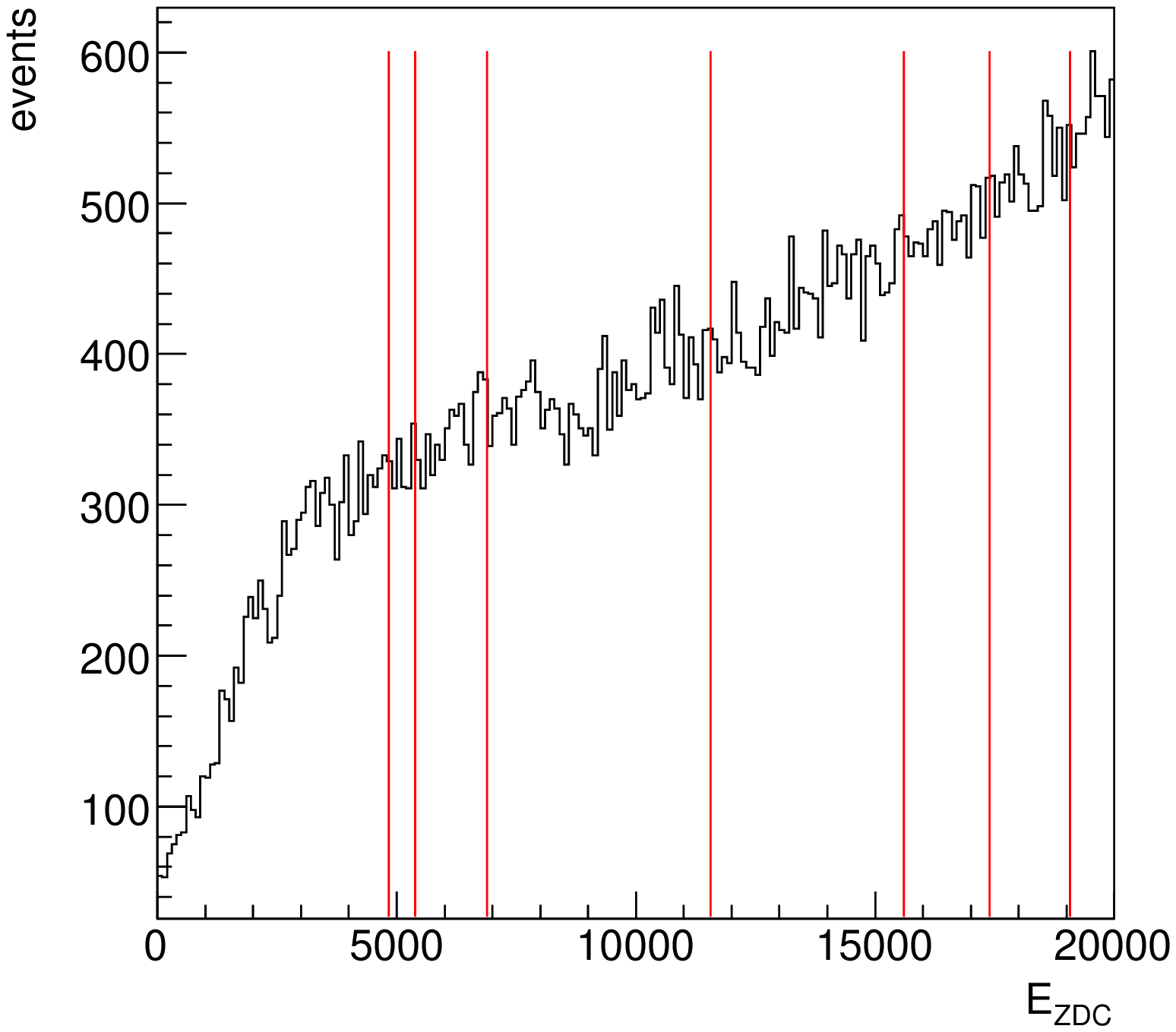}
\includegraphics[width=0.49\textwidth]{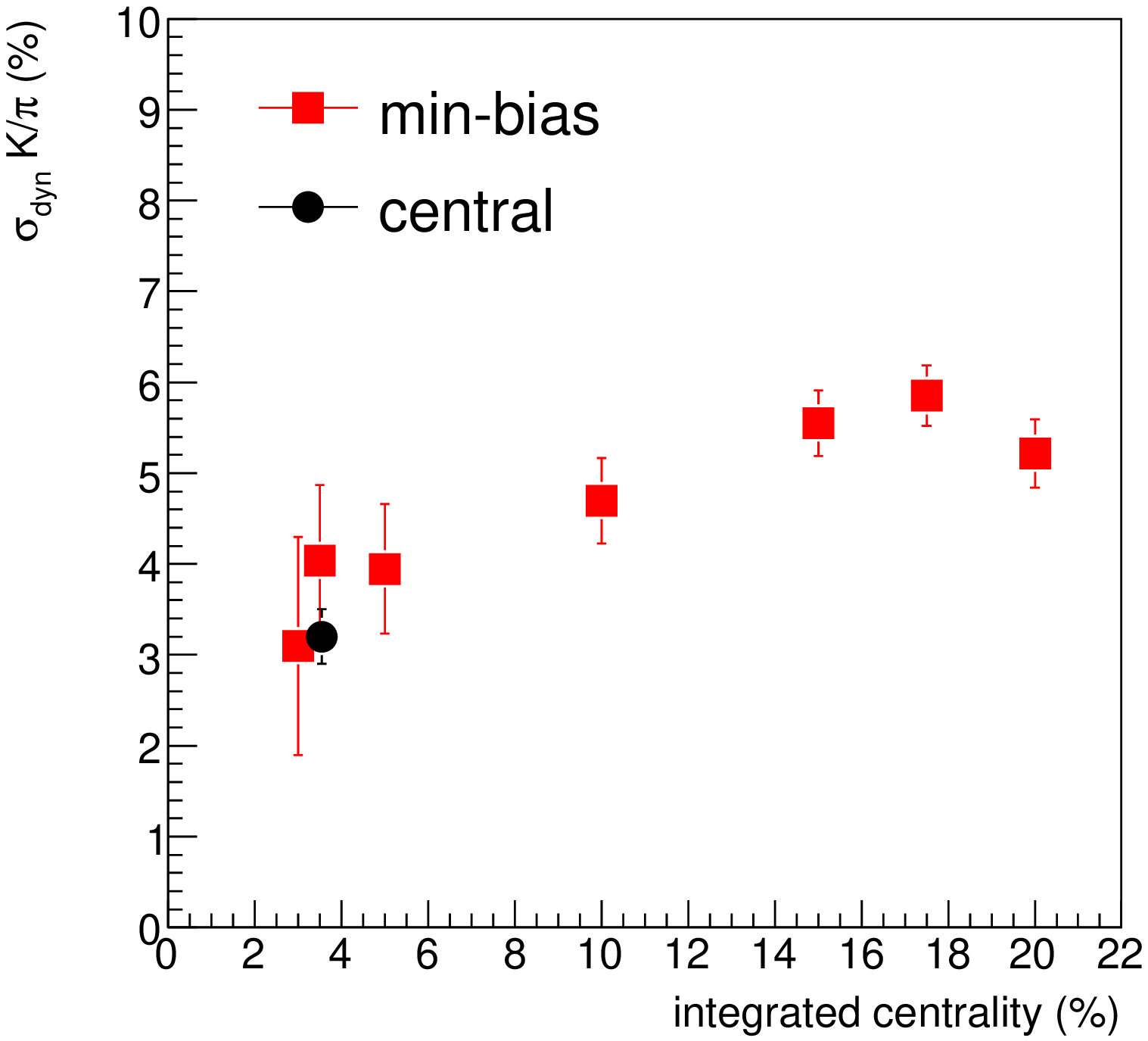}
\caption{Left picture: distribution of the energy deposited in the Zero
Degree Calorimeter (ZDC) of NA49 in Pb + Pb collisions at
158$A$~GeV beam energy. Vertical lines show cuts, which correspond
to 3\%, 3.5\%, 5\%, 10\%, 15\%, 17.5\% and 20\% centrality bin size.
Right picture: dynamical fluctuations of the K/$\pi$ ratio
as a function of the centrality bin size in Pb + Pb collisions
at 158$A$~GeV beam energy.}
\label{fig:na49:cbins}
\end{figure}

The main motivation to use particle yield ratios in the fluctuation
analysis is that in first approximation volume fluctuations are cancelled in
the ratio. Nevertheless, the fluctuations of the projectile and target
participants in Pb + Pb collisions may have an influence on the measured signal,
thus more detailed studies had been performed on the subject
of the dependence of dynamical fluctuations on the selected centrality
bin size. In the published NA49 results on the energy dependence
of the 3.5\% most central Pb + Pb collisions were used in the
analysis~\cite{:2008ca,bib:croland}. \\
The distribution of the total energy of the projectile spectators deposited in
the Zero Degree Calori-meter of NA49 is shown in figure~\ref{fig:na49:cbins}
(left plot). The centrality was integrated in the following bins: (0-3)\%,
(0-3.5)\%, (0-5)\%, ..., (0-20)\% most central Pb + Pb collisions at
158$A$~GeV beam energy. The observed dependence of the K/$\pi$ ratio fluctuations
is shown in figure~\ref{fig:na49:cbins} (right plot).

The results are in good agreement with the top SPS beam energy point
from~\cite{:2008ca}. As has been discussed before, the fluctuations
of the number of projectile and target participants
have an effect on the measured signal, and a slight increase of about
2\% of fluctuations with increasing integrated centrality of Pb + Pb collisions
is observed as a general trend. However, the difference between the values of the
dynamical fluctuations of the kaon to pion yield ratio in case of using
3.5\% and 5\% most central Pb + Pb collisions is small. So, in order to gain
statistics, which is important for the errors, the centrality dependence
of the dynamical fluctuations can be studied using a centrality bin size
of 5\%.

\subsection{Simulations with UrQMD}

\begin{figure}[ht]
\center
\includegraphics[width=0.5\textwidth]{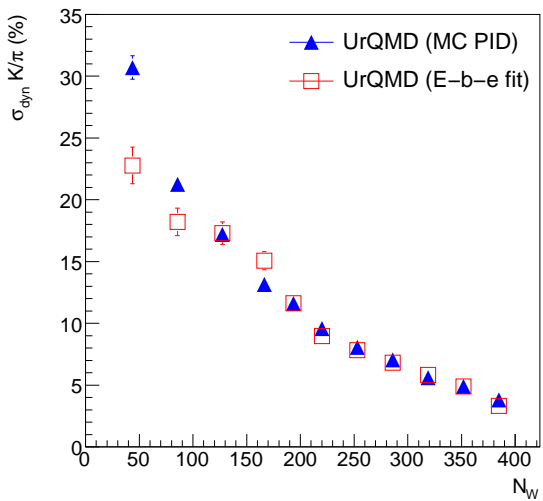}
\caption{Dynamical fluctuations of the K/$\pi$ ratio
as a function of the number of wounded nucleons in Pb + Pb collisions
at 158$A$~GeV beam energy, simulated with UrQMD, for two cases:
Monte Carlo based counting (triangles) and event-by-event fit
based on simulated $dE/dx$ response (squares).}
\label{fig:na49:urqmd:diff}
\end{figure}

Particle identification on the statistical basis can introduce a bias to the
event-by-event fluctuations. This bias might be larger in semi-peripheral
Pb + Pb collisions, since in this case the average track multiplicity
decreases and the identification procedure becomes more difficult.
It becomes particularly challenging if single particle multiplicities in the
acceptance approach zero. As no negative numbers can be allowed for stable
fit results, the eventwise K/$\pi$ distribution develops a spike at
zero~\cite{:2008ca,bib:dkresan}. \\
In order to find out the working limits for the analysis of the centrality
dependence of Pb + Pb collisions, simulations with the UrQMD model~\cite{Bass:1998ca}
have been performed. The centrality of a collision was determined using the provided
value of the impact parameter. Particles, generated by the model, were
processed through the acceptance filter of NA49, which incorporates the specific
track cuts used in the current data analysis. For each accepted track, a $dE/dx$ value
was simulated using a parametrization of the measured $dE/dx$ distribution in NA49.
The accepted tracks with their simulated $dE/dx$ response were processed through
the same analysis routines as the tracks from data, namely the event-by-event fit of
the eventwise $dE/dx$ distribution (E-b-e fit). As an unbiased reference, particle identification
based on the particle type provided by the UrQMD model was used, so called Monte Carlo
identification (MC PID). The extracted centrality dependence of the dynamical
fluctuations of the kaon to pion yield ratio for these two cases is shown
in figure~\ref{fig:na49:urqmd:diff}.

For both identification procedures the dynamical fluctuations increase
with decreasing number of wounded nucleons. But the results for the event-by-event
fit start to deviate from the MC reference for less than 200 wounded nucleons,
which approximately corresponds to a centrality of 35\%. The contribution
of the spike at zero in this case exceeds 2\% of all measured events.
Thus NA49 data will be analyzed in the centrality range from 0\% to
35\% most central Pb + Pb collisions, where the identification procedure does not
introduce a bias to the results.

\subsection{Centrality Dependence}

\begin{figure}[ht]
\center
\includegraphics[width=0.45\textwidth]{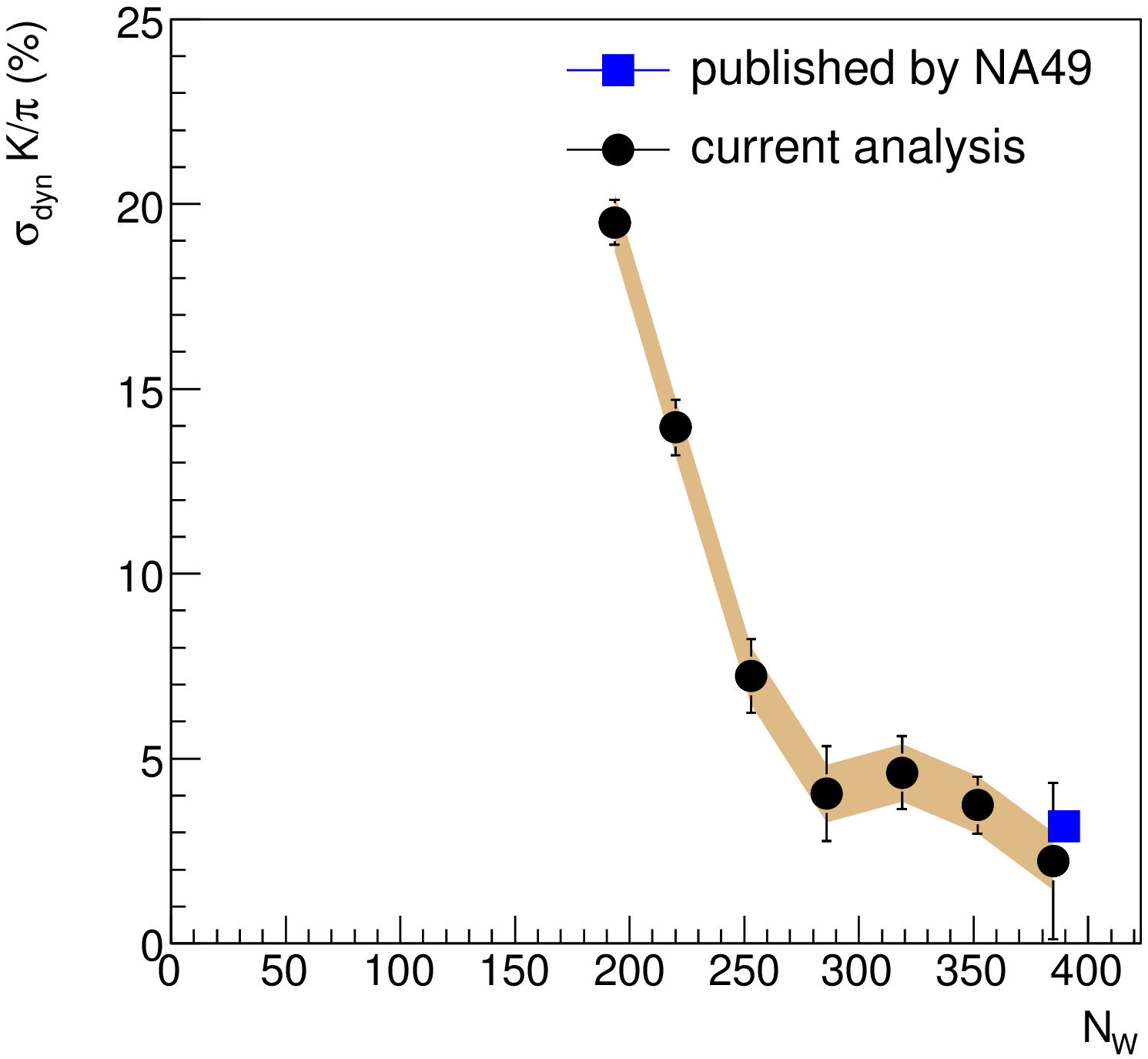}
\includegraphics[width=0.45\textwidth]{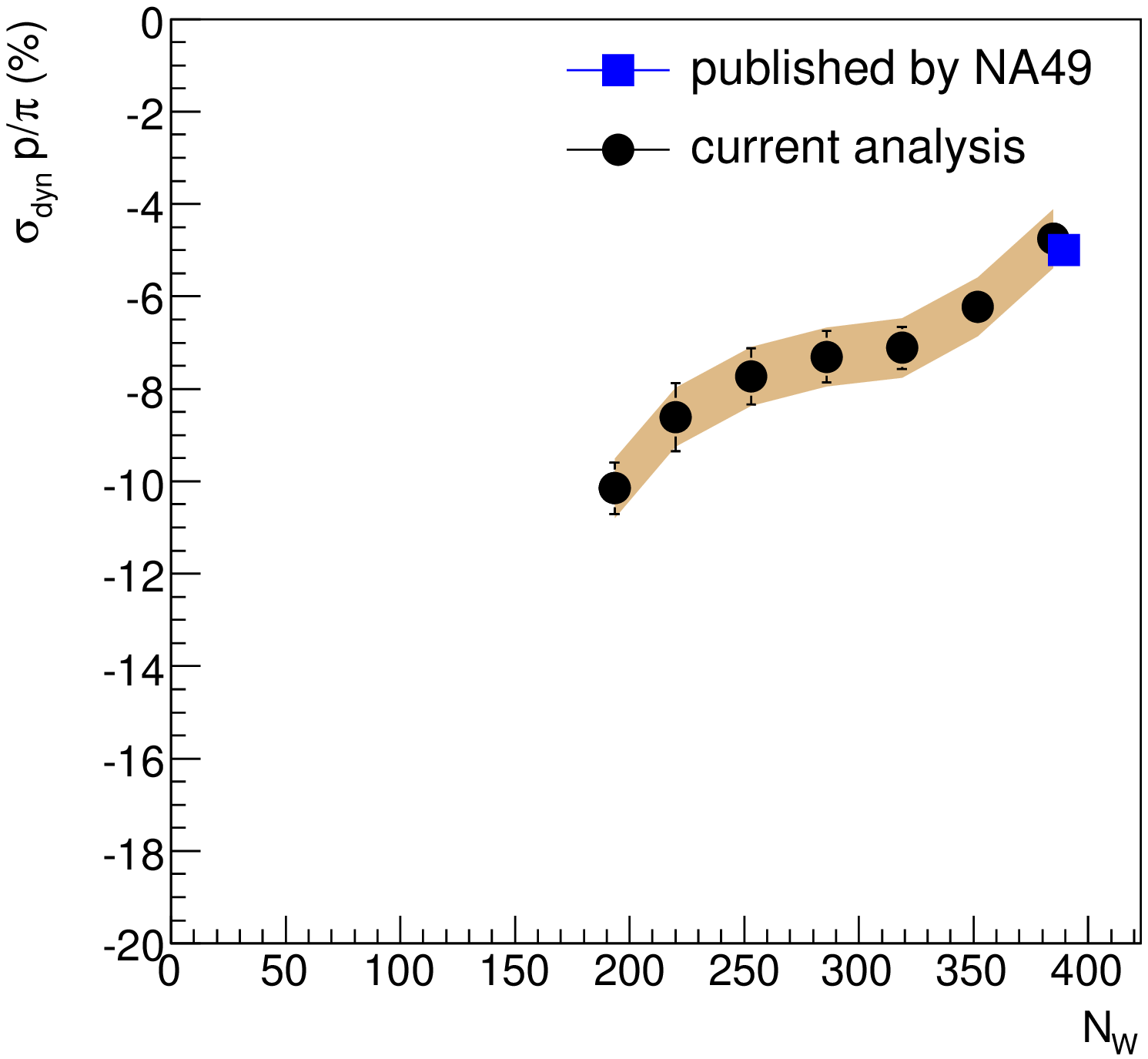}
\caption{Dynamical fluctuations of the K/$\pi$ (left picture) and the
p/$\pi$ (right picture) ratios as a function of the number of wounded
nucleons in Pb + Pb collisions at 158$A$~GeV beam energy. The shaded band
shows systematic errors determined from a variation of track cuts.}
\label{fig:na49:diff:1}
\end{figure}

The measured centrality dependences of the dynamical fluctuations of the
K/$\pi$ and p/$\pi$ ratios in Pb + Pb collisions at 158$A$~GeV beam energy are
presented in figure~\ref{fig:na49:diff:1}.

\begin{figure}[ht]
\center
\includegraphics[width=0.45\textwidth]{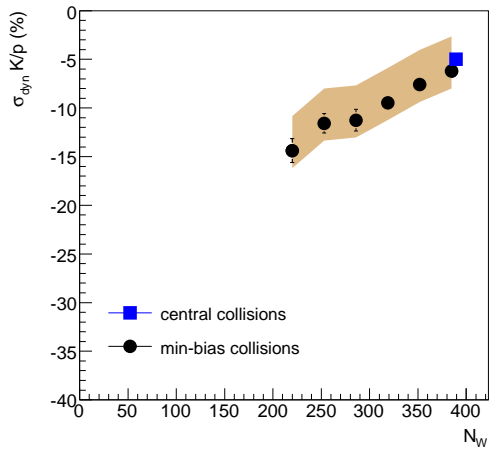}
\caption{Dynamical fluctuations of the K/p ratio as a function
of number of wounded nucleons in Pb + Pb collisions at 158$A$~GeV beam energy.
The shaded band shows systematic errors determined from a variation of track cuts.}
\label{fig:na49:diff:2}
\end{figure}

Figure~\ref{fig:na49:diff:2} shows the dynamical fluctuations of the K/p ratio
as a function of the number of wounded nucleons in Pb + Pb collisions at 158$A$~GeV
beam energy. The fluctuations are calculated as mean value from analysis with
different track cuts. Their difference is used to estimate the systematic error
which is given by the shaded band.

The general trend is the same for all considered particle yield ratio fluctuations:
the absolute value of the dynamical fluctuations increases towards lower centrality
of Pb + Pb collisions.

\subsection{Scaling of Dynamical Fluctuations}

From the definition of the dynamical fluctuations (equation~ \ref{eq:na49:sdyn})
one can derive an exact analytical expression, which in case of K/$\pi$ consists of kaon,
pion and correlation terms:

\begin{equation}
\sigma_{dyn} = \sqrt{\frac{var(N_{K})}{<N_{K}>^{2}} + \frac{var(N_{\pi})}{<N_{\pi}>^{2}} - 2\frac{cov(N_{K},N_{\pi})}{<N_{K}><N_{\pi}>}},
\label{eq:na49:sdyn:full}
\end{equation}

In case of p/$\pi$, $N_{K}$ should be substituted with $N_{p}$.
As has been argued in~\cite{:2008ca,Kresan:2006zz}, dynamical fluctuations of the p/$\pi$ 
ratio originate dominantly from $\Delta$ resonance decays, which produces
correlated protons and pions. The analytical expression for the dynamical
fluctuations in this case can be approximated by the correlation term only:

\begin{equation}
\sigma_{dyn}^{p/\pi} \approx -\sqrt{\frac{cov(N_{p}, N_{\pi})}{<N_{p}><N_{\pi}>}} = -\sqrt{\frac{(<N_{p}><N_{\pi}>)^{\alpha}}{<N_{p}><N_{\pi}>}},
\label{eq:na49:prot2pi}
\end{equation}

where we have parametrised the covariance by the product of average multiplicities
of protons and pions
to the power of $\alpha$. With a strong feeddown from resonances this parameter $\alpha$
is expected to become $0.5$. In case of the K/$\pi$ ratio we have
$N_{K} \ll N_{\pi}$ and the analytical expression for $\sigma_{dyn}$ can be
apporoximated by:

\begin{equation}
\sigma_{dyn}^{K/\pi} \approx \sqrt{\frac{var(N_{K})}{<N_{K}>^{2}}}.
\end{equation}

\begin{figure}[ht]
\includegraphics[width=0.49\textwidth]{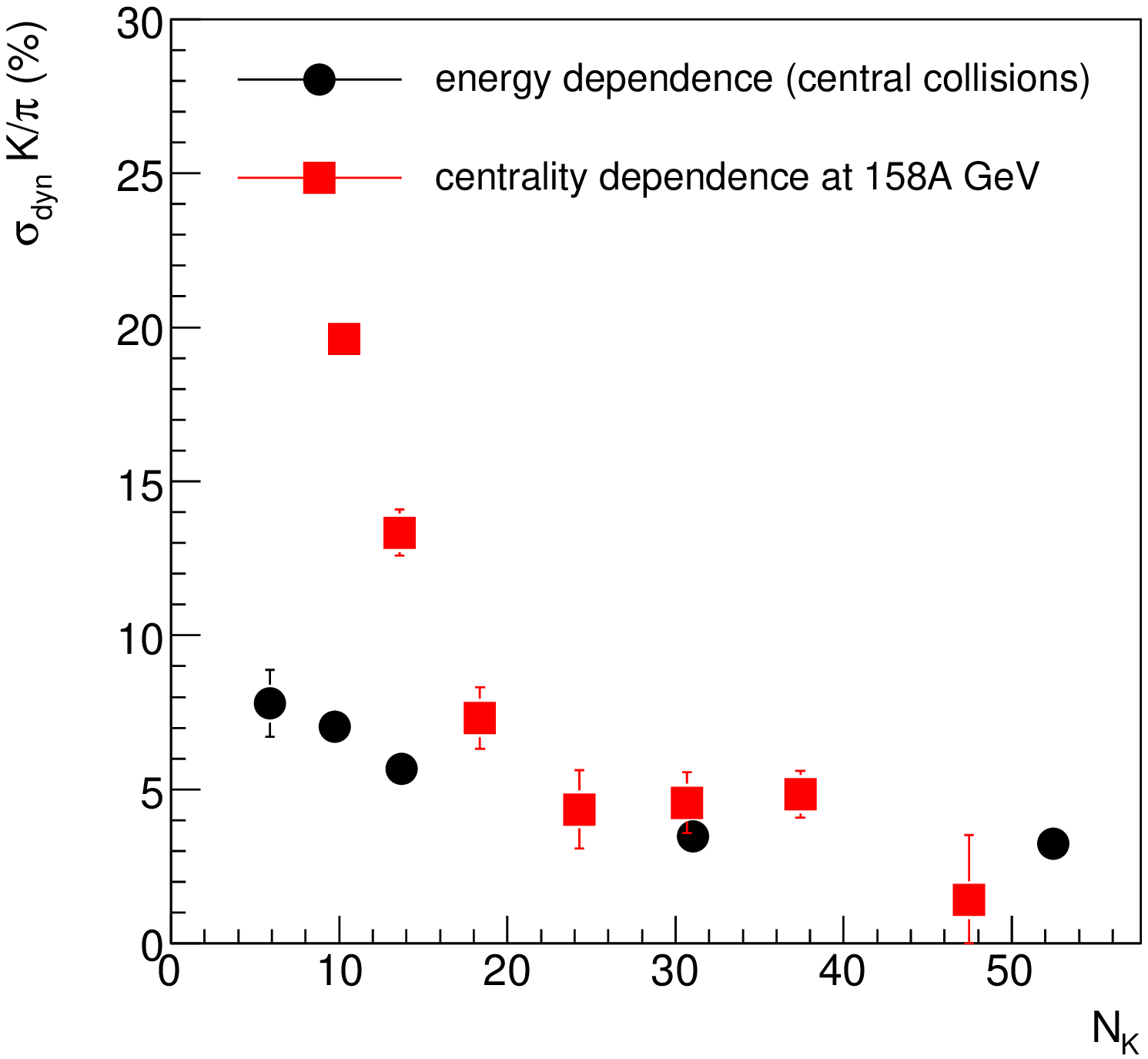}
\includegraphics[width=0.49\textwidth]{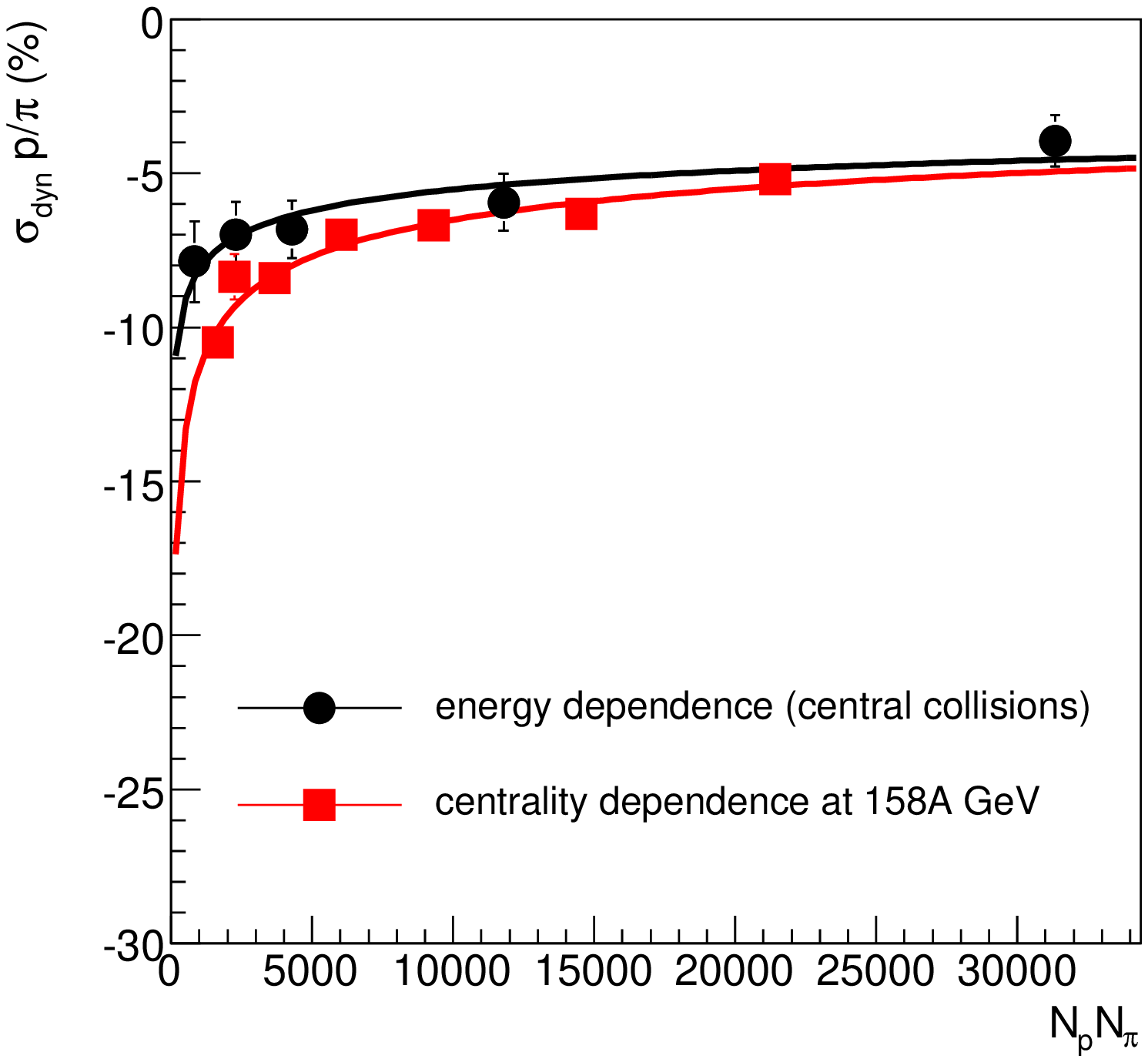}
\caption{Left picture: dynamical fluctuations of the K/$\pi$ ratio
as a function of $<N_{K}>$ for
the energy (circles) and the centrality (squares) dependences.
Right picture: dynamical fluctuations of the p/$\pi$ ratio
as a function of $<N_{p}><N_{\pi}>$
for the energy (circles) and the centrality (squares) dependences.
Solid curves show fits according to equation~2.3.}
\label{fig:na49:scaling}
\end{figure}

Figure~\ref{fig:na49:scaling} shows the dynamical fluctuations of the K/$\pi$
ratio as a function of $<N_{K}>$ (left picture) and
the dynamical fluctuations of the p/$\pi$ ratio as a function of
$<N_{p}><N_{\pi}>$ (right picture). Data points
are taken from the measured energy and centrality dependences for a more
loose set of track cuts; the mean number
of particles is calculated in the acceptance used to determine the fluctuations.
After fitting the dependences of the p/$\pi$ ratio fluctuations with
equation~\ref{eq:na49:prot2pi}, the following $\alpha$ parameters were obtained:
$\alpha=0.66\pm0.12$ for the energy dependence and $\alpha=0.51\pm0.03$
for the centrality dependence. This observation strongly supports the assumption
that $\Delta$ resonance decays are the dominant source of dynamical fluctuations
of the p/$\pi$ ratio and give a natural explanation for both, the energy
and centrality dependence. As expected also for K/$\pi$ ratio fluctuations
some dependence on $<N_{K}>$ is seen, however the energy and centrality
dependence obviously scale differently with $<N_{K}>$. Whether this might
be connected to different relative strangeness production in smaller
systems~\cite{Hohne:2005ks} requires further studies.

\section{Particle Ratio Fluctuations in CBM}

\begin{figure}[ht]
\center
\includegraphics[width=0.5\textwidth]{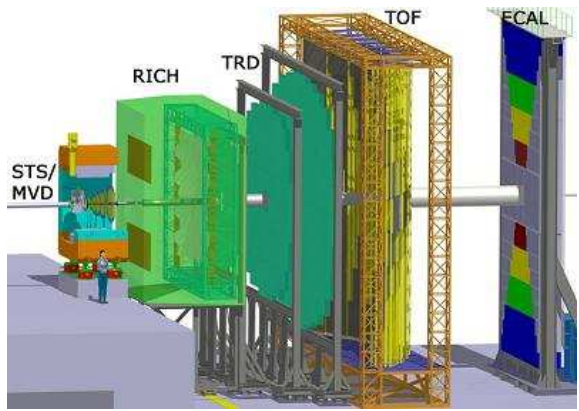}
\caption{Sketch of the CBM detector.}
\label{fig:cbm:det}
\end{figure}

CBM is the future heavy-ion fixed target experiment at FAIR~\cite{bib:cbm:tsr}. Its goal
is to measure hadrons and leptons in A + A collisions at beam energies from 10 - 45~AGeV.
The proposed CBM detector (electron option) is shown in figure~\ref{fig:cbm:det}.

\begin{figure}[ht]
\center
\includegraphics[width=0.9\textwidth]{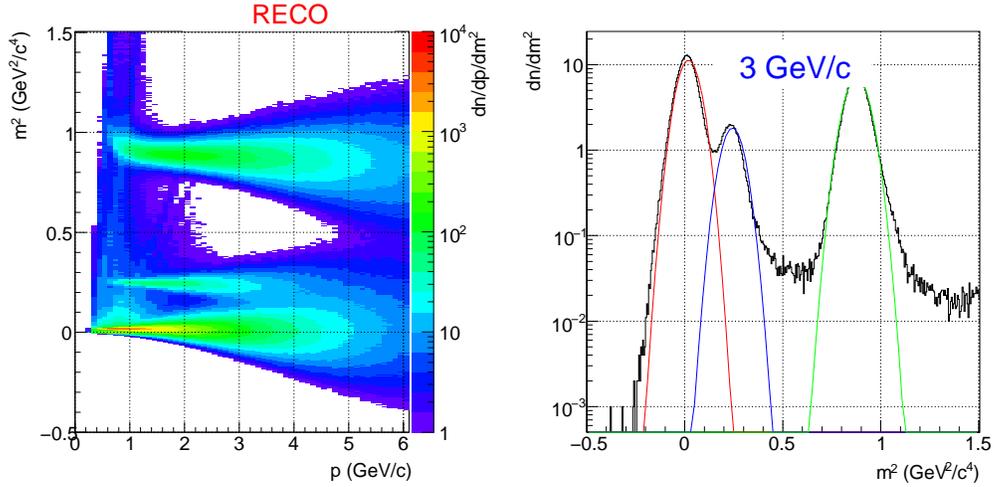}
\caption{Distribution of the squared mass versus momentum of reconstructed
primary hadrons from central Au + Au collisions at 25$A$~GeV beam energy (UrQMD simulation).
Right picture shows a projection for the momentum of 3~GeV/c. The tails originate
from secondary particles and ghost tracks (mismatches).}
\label{fig:cbm:m2mom}
\end{figure}

In this section we will focus on feasibility studies of the measurement of
event-by-event fluctuations of the K/$\pi$ ratio with CBM.
In CBM hadrons will be identified using a TOF wall. With a time resolution
of better than 80~ps it will be possible to identify particles on the
track-by-track level, which together with a large acceptance (larger as compared
to the NA49 experiment) should avoid the development of a spike at zero in the eventwise
K/$\pi$ ratio distributions. The distribution of the squared mass versus momentum of
reconstructed primary hadrons from UrQMD~\cite{Bass:1998ca} simulations
is shown in figure~\ref{fig:cbm:m2mom}.

A clean identification of kaons with a pion contamination on the level of 15\% only
(integrated) is possible up to momenta of approximately 3~GeV/c.

\begin{figure}[ht]
\center
\includegraphics[width=0.47\textwidth]{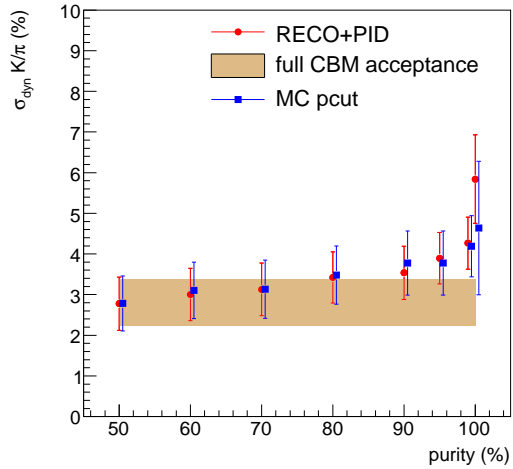}
\caption{Dynamical fluctuations of the K/$\pi$ ratio as a function
of purity of the kaon identification for the standard identification procedure
(circles), Monte Carlo based counting with the upper momentum cut, which
corresponds to the chosen purity, (squares) and full CBM acceptance
(filled region).}
\label{fig:cbm:purity}
\end{figure}

A certain level of the purity of kaon identification is achieved by using
an upper momentum cut. As this of course restricts the phase space acceptance,
the effect of this cut on the dynamical fluctuations of the K/$\pi$ ratio was investigated.
As a reference, the results with Monte Carlo (MC) identification but the same
upper momentum cut were calculated. The observed dependence
of the dynamical fluctuations on the purity of kaon identification i.e. the applied
momentum cut is presented in figure~\ref{fig:cbm:purity} for MC identification
and fully reconstructed and identified tracks in the CBM simulation.

For example the purity requirement of at least 50\% per momentum bin corresponds
to a momentum cut of 6~GeV/c, 99\% to 3.25~GeV/c.
Values of the dynamical fluctuations increase if the momentum cut is more restrictive.
Good agreement between results with real identification and PID on the MC level
with corresponding momentum cut is observed. The results are closer to the values for
the full CBM acceptance in case of lower purity, which corresponds to a more relaxed
total momentum cut.

\section{Summary and Outlook}

The centrality dependences of K/$\pi$, p/$\pi$ and K/p ratio
fluctuations show increasing absolute values towards lower centrality. A
multiplicity dependence is indeed expected and the dependence of the p/$\pi$
ratio fluctuations on energy and centrality can both together be well described
assuming a dominant contribution from $\Delta$ resonance decays. In case of
the K/$\pi$ ratio fluctuations however, the values of the dynamical
fluctuations of the K/$\pi$ ratio for semi-peripheral Pb + Pb
collisions at 158$A$ GeV are higher than those for central Pb + Pb collisions
at lower SPS energies at the same values of $<N_{K}>$.
In other words: the energy and centrality dependences of the K/$\pi$ ratio
fluctuations scale differently with $<N_{K}>$. \\
As dynamical fluctuations in particle emmision are expected to be an important
observable for the critical endpoint and current data are not conclusive yet,
the future CBM experiment at FAIR prepares to also provide precise
evaluation/observation of these fluctuations. Feasibility studies for the measurement of
event-by-event fluctuations of the K/$\pi$ ratio in central Au + Au collisions
were presented and demonststrate that some of the experimental difficulties in the NA49
analysis can be overcome. The study shows that such a measurement is feasible
and that no strong bias from the identification procedure with the TOF wall
of CBM is expected.




\begin{thebibliography}{99}

\bibitem{Fodor:2004nz}
Z.~Fodor and S.~D.~Katz,
JHEP {\bf 0404} (2004) 050
[arXiv:hep-lat/0402006].

\bibitem{Aoki:2004iq}
S.~Aoki {\it et al.}  [JLQCD Collaboration],
Phys.\ Rev.\  D {\bf 72} (2005) 054510
[arXiv:hep-lat/0409016].

\bibitem{Ejiri:2007ga}
S.~Ejiri,
Phys.\ Rev.\  D {\bf 77} (2008) 014508
[arXiv:0706.3549 [hep-lat]].

\bibitem{Allton:2003vx}
C.~R.~Allton, S.~Ejiri, S.~J.~Hands, O.~Kaczmarek, F.~Karsch, E.~Laermann and C.~Schmidt,
Phys.\ Rev.\  D {\bf 68} (2003) 014507
[arXiv:hep-lat/0305007].

\bibitem{Koch:2008ia}
V.~Koch,
arXiv:0810.2520 [nucl-th].

\bibitem{:2008ca}
C.~Alt {\it et al.}  [NA49 Collaboration],
arXiv:0808.1237 [nucl-ex].

\bibitem{:2009if}
B.~I.~Abelev {\it et al.}  [STAR Collaboration],
arXiv:0901.1795 [nucl-ex].

\bibitem{Afanasev:1999iu}
S.~Afanasev {\it et al.}  [NA49 Collaboration],
Nucl.\ Instrum.\ Meth.\  A {\bf 430} (1999) 210.

\bibitem{Wenig:1998vv}
S.~Wenig  [NA49 Collaboration],
Nucl.\ Instrum.\ Meth.\  A {\bf 409} (1998) 100.

\bibitem{bib:croland}
C.~Roland, Ph.D. thesis, J.~W.~G.~Universit\"at, Frankfurt am Main, (1999).

\bibitem{bib:dkresan}
D.~Kresan, Ph.D. thesis, J.~W.~G.~Universit\"at, Frankfurt am Main,
to be submitted in 2009.

\bibitem{Bass:1998ca}
S.~A.~Bass {\it et al.},
Prog.\ Part.\ Nucl.\ Phys.\  {\bf 41} (1998) 255
[Prog.\ Part.\ Nucl.\ Phys.\  {\bf 41} (1998) 225]
[arXiv:nucl-th/9803035].

\bibitem{Kresan:2006zz}
D.~Kresan and V.~Friese,
PoS C {\bf FRNC2006} (2006) 017.

\bibitem{Hohne:2005ks}
C.~Hohne, F.~Puhlhofer and R.~Stock,
Phys.\ Lett.\  B {\bf 640} (2006) 96
[arXiv:hep-ph/0507276].

\bibitem{bib:cbm:tsr} CBM Collaboration,
http://www.gsi.de/onTEAM/dokumente/public/DOC-2005-Feb-447.html.

\end{thebibliography}
\end{document}